\documentclass[floatfix, twoside, twocolumn, aps]{revtex4}
\usepackage{epsfig}

\begin{document}

\title{Predicting Multivariate Volatility}

\author{Christof Reese and Bernd Rosenow}

\affiliation{Institut f\"ur Theoretische Physik, Universit\"at zu
K\"oln, D-50923 Germany}

\date{\today}

\begin{abstract}
  We suggest two classes of multivariate GARCH--models which are both
  easy to estimate and perform well in forecasting the covariance
  matrix of more than one hundred stocks. We apply methods from random
  matrix theory (RMT) to determine the number of principal components
  or the number of factors in the multivariate volatility models.  In
  this way only statistically relevant information is used for the
  estimation of model parameters.
\end{abstract}

\maketitle

The basic input for the estimation of portfolio risk is a forecast of
time varying covariance matrices.  Although modern risk management is
founded on the analysis of huge data bases and the use of
sophisticated theoretical models, there has been only partial progress
with respect to covariance forecasts.  The reason for this lack of
progress is the ``curse of dimensionality'': in modern financial
engineering, the dimension of the investment universe is comparable to
the number of observations, and hence the number of parameters in the
covariance or correlation matrix is of the same order as the number of
data points.

Traditional multivariate statistics is not able to deal with this kind
of problem. Both estimators and tests are consistent only in the limit
of $L/N$ going to infinity, where $L$ is the number of observations
and $N$ is the dimension of the observed vector
\cite{Muirhead82}.  In typical applications, one is dealing
with ratios $L/N$ ranging from one to ten, and hence statistical
fluctuations due to the lack of a sufficient number of observations
dominate an empirically observed covariance matrix. This measurement
noise induces randomness in the covariances and hence turns the
covariance matrix at least partly in a random matrix.

A covariance matrix can be decomposed into the cross--correlation
matrix {\bf \sf C} and the set of standard deviations
$\{\sigma_i^2\}$. To identify the effects of randomness on the
eigenvalue spectrum of an empirical cross--correlation matrix, one
considers the null hypothesis of completely uncorrelated and
identically distributed (i.i.d.) time series.  The cross--correlation
matrix calculated from these time series is a random matrix {\bf \sf
  R} from the so called Wishart ensemble. The eigenvalues of a random
Wishart matrix fall in the interval $[ \lambda_-, \lambda_+]$ with
$\lambda_\pm = 1 + N/L + 2 \sqrt{N/L}$ \cite{Marcenko+67}. This
eigenvalue distribution can be compared to that of the empirical
cross--correlation matrix {\bf \sf C}.  Agreement between {\bf \sf R}
and {\bf \sf C} is a sign of randomness, whereas deviations indicate
the presence of economically meaningful information. It was found that
approximately 94 \% of the empirically observed eigenvalues fall in
the random matrix interval \cite{Laloux+99}. The random matrix nature
of the main part of the spectrum was proven by comparing the
eigenvalue statistics of empirical correlation matrices to the
universal predictions of random matrix theory \cite{Plerou+99}.  Only
the largest eigenvalues and corresponding eigenvectors carry
information about market correlations, business sectors and
geographical correlations \cite{Gopikrishnan+01,Plerou+2002}. These
economically meaningful correlations are stable in time and hence
suitable for forecasting future correlations.  Generally, correlations
show a higher degree of time stability than covariances which are
fluctuating due to time changing volatility.

The time evolution of the variance or volatility is often described
with GARCH--models \cite{Engle95}, which relate the present variance
to past variances and past returns.  While GARCH--processes are well
suited to describe the dynamics of a univariate volatility process
\cite{Anderson97}, they are less successful in the multivariate
setting.  In addition to the ``curse of dimensionality'', the
application of multivariate GARCH--models is hampered by the need to
make maximum likelihood estimates of a large number of parameters
growing generically like the square of the number of time series.  In
this note, we suggest two classes of multivariate GARCH--models, which
incorporate RMT based progress in the understanding of empirical
cross--correlation matrices. These models are both easy to estimate
and compare favorably to alternative models as the RiskMetrics
covariance estimator \cite{Longerstaey+96}.\\

\noindent
{\bf High dimensional correlation matrices}

\begin{figure}
\centerline{ \epsfig{file=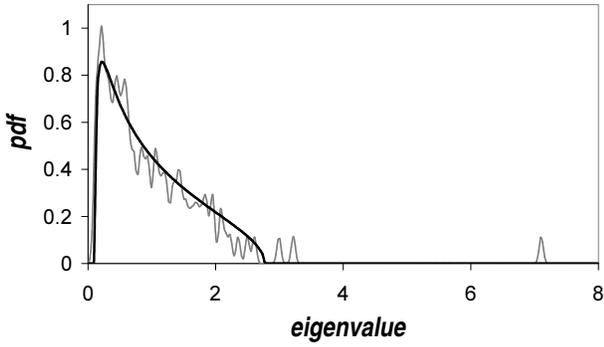,width=8cm} }
\caption{ Probability distribution of the eigenvalues  of the
cross--correlation matrix {\bf \sf C}$^{\rm a}$ calculated from daily
118 return time series from 09/01/99 until 08/31/00 (gray line) . The
bulk of eigenvalues is due to noise and well described by the
prediction of random matrix theory (black line).  One eigenvalue is
clearly separated from the bulk and contains information about market
correlations.}
\label{spektrum}
\end{figure}

\noindent
We estimate the sample covariance matrix
%
\begin{equation}
\Sigma_{ij}= {1 \over (L -1)} \sum_{t=1}^L (\epsilon_{i,t} - \langle
\epsilon_i\rangle)(\epsilon_{j,t} - \langle \epsilon_j \rangle)
\end{equation}
%
and define the sample correlation matrix $C_{ij}=\Sigma_{ij}/(\sigma_i
\sigma_j)$ by normalizing with the standard deviations $\sigma_i =
\sqrt{\Sigma_{ii}}$. We calculate cross--correlation matrices for the
118 most frequently traded German companies from a) 1711
daily returns $\epsilon_{i,t}$ starting 12/01/93 until 08/31/00, and
b) 250 daily returns starting 09/01/99 until 08/31/00, diagonalize
them, and rank-order their eigenvalues $\lambda_i < \lambda_k$ for
$i<k$.  The pdf of the eigenvalues of {\bf \sf C}$_{\rm a}$ is
displayed in  Fig. \ref{spektrum}. For both {\bf \sf C}$_{\rm
a}$ and {\bf \sf C}${\rm _b}$ we find that three eigenvalues lie
outside the random matrix interval, but only one of them clearly. The
number of eigenvalues clearly outside the RMT--interval is the optimal
number of principal components or factors that should be used to
forecast the correlation matrix \cite{Rosenow2002}.

The largest eigenvalue $\lambda_{\rm max}=14.5$ for the
seven--year interval is much larger than $\lambda_{\rm max}=7.1$
for the one--year interval. This confirms the result of
\cite{Plerou+99,drozdz00,Plerou+2002} that the strength of market correlations changes
in time.  Although the use of time series extending over several
years in theory increases the accuracy of correlation estimates,
it does not capture the effect of time changing correlations and
hence is not advisable.

In order to use only the statistically relevant information in the
correlation matrix {\bf \sf C}, we diagonalize {\bf \sf C} by an
orthogonal transformation ${\bf \sf X}$ via ${\bf \sf X}^T {\bf \sf C}
{\bf \sf X}= {\bf \sf \Lambda}$. We construct a filtered correlation
matrix ${\bf \sf C}_p^\prime$ by keeping only the $p$ largest
eigenvalues in the diagonal matrix \cite{Gopikrishnan+01,Laloux+99,Rosenow2002}
%
\begin{equation}
\Lambda^{\rm p}_{ii}= \left\{ \begin{array}{l} \Lambda_{ii} \ \ {\rm for}\
i > N - p \\
0 \ \ {\rm for } \ i \leq N - p \end{array} \right. \ ,
\end{equation}
%
and by transforming back to the original basis ${\bf \sf C}^{\rm p} =
{\bf \sf X} {\bf \sf \Lambda}^{\rm p} {\bf \sf X}^T$. For consistency,
all
diagonal elements of ${\bf \sf C}^{\rm p}$ are set to one.\\

\noindent
{\bf Model description}

\noindent

The basis of the {\it Sliding Correlation Model} (SCM) is the
decomposition of the covariance matrix into the cross--correlation
matrix {\bf \sf C} and standard deviations $\{\sigma_i\}$.  Each
individual time series is described by a univariate GARCH(1,1)
\cite{Engle95} process defined by
%
\begin{equation}
    \sigma_{i,t}^2=\alpha_{i,0}+\alpha_{i,1}\sigma_{i,t-1}^2+
\beta_{i,1}\epsilon_{i,t-1}^2 \ .
\label{garch}
\end{equation}
%
Here, $\sigma_{i,t}^2$ and $\epsilon_{i,t}$ are the volatility and the
innovation of time series $i$ at time t, respectively. The innovations
are modelled as a product $\epsilon_{i,t} = \sigma_{i,t} \cdot x_{i,t}$
with normally distributed $x \in {\cal N}(0,1)$. The parameters
$\alpha_{i,0}$, $\alpha_{i,1}$, and $\beta_{i,1}$ are obtained from
maximum likelihood estimations.  The cross--correlation matrix {\bf
\sf C} is calculated as a moving average to
accommodate the change of correlations in time.  We use the filtering
method described in the last section and keep only the $p$ largest
principal components.  The covariance matrix in the sliding
correlation model SCM(p) of rank $p$ is described by
%
\begin{equation}
    \Sigma_{ij,t}=C^{\rm p}_{ij,t} \sigma_{i,t} \ \sigma_{j,t} \ .
\end{equation}
%
This model is related to the constant correlation model
\cite{Bollerslev90}, where a fixed and unfiltered correlation matrix
is used in contrast to our $C^{\rm p}_{ij,t}$. 

{\it Factor Models} are motivated by the observation that correlations
are large in times of high volatility \cite{Plerou+2002}. In addition,
general multivariate GARCH--models can be simplified to factor models,
if the parameter matrices do not have full rank.  As a special case
with parameter matrices of rank one, a one--factor model was discussed
in \cite{Engle1990}.  We suggest a more general model with $p$ factors
(abbreviated Factor(p))
%
\begin{equation}
    \Sigma_{ij,t}=\tilde{\sigma}_{i,t}^2\delta_{ij}+\sum_{k=1}^p
    \lambda_i^{(k)}
\ \lambda_j^{(k)} \ \Sigma^{(k)}_{t} \ .
\end{equation}
%
Here, $\tilde{\sigma}_{i,t}^2$ is the variance of the residual of time
series $i$ after linear regression on the $p$ factors $f^{(k)}_t$ with
$k=1 ... p$, and $\lambda^{(k)}_i$ is the regression coefficient
(factor loading) of factor $k$ on time series $i$.  The dynamics of
the residuals is described by a GARCH(1,1)--process, and the
dynamics of the factors is similarly given by
%
\begin{equation}
  \Sigma^{(k)}_t
= \alpha_0^{(k)} + \beta_1^{(k)} \Sigma^{(k)}_{t-1} + \alpha_1^{(k)}
(f_{t-1}^{(k)})^2 \ .
\end{equation}
%
The factors $f_t^{(k)}$ are defined via an iterative procedure: we
use the largest eigenvector ${\bf u}^{(N)}$ of ${\bf \sf C}$ to
define $f^{(1)}_t = \sum_{i=1}^N u^{(N)}_i \epsilon_{i,t}$. Next,
we perform a linear regression of $f^{(1)}$ on the $\epsilon_i$
and recompute the correlation matrix from the residuals. From the
new eigenvector corresponding to the largest eigenvalue, we define
a factor $f^{(2)}$, perform another linear regression and so on.
The number $p$ of factors is chosen according to the number of
eigenvalues of the correlation matrix outside the 
RMT--interval, the time interval for the calculation of factors
and factor loadings is not moving but fixed.

The data set for parameter estimation comprises seven years of daily
data from 12/01/93 until 08/31/00, which are used for the maximum
likelihood estimation of all GARCH parameters, for the calculation
of factors and factor loadings of the models Factor(p)-seven, whereas
only the last year of daily data is used for models Factor(p)-one.
Similarly, the cross--correlation matrix of model SCM(p)-seven is
calculated over a sliding time window of length seven year, and a
window of length one year is used for model SCM(p)-one.

Our benchmark model is the commonly used  Risk Metrics 
covariance estimator \cite{Longerstaey+96} defined by
%
\begin{equation}
    \Sigma_{ij,t}\ =\ 0.94 \ \Sigma_{ij,t-1}\ +\ 0.06 \ \epsilon_{i,t}\,
 \epsilon_{j,t}\ .
\label{riskmetrics.eq}
\end{equation}\\[-.2cm]
%

\begin{figure}
\centerline{ \epsfig{file=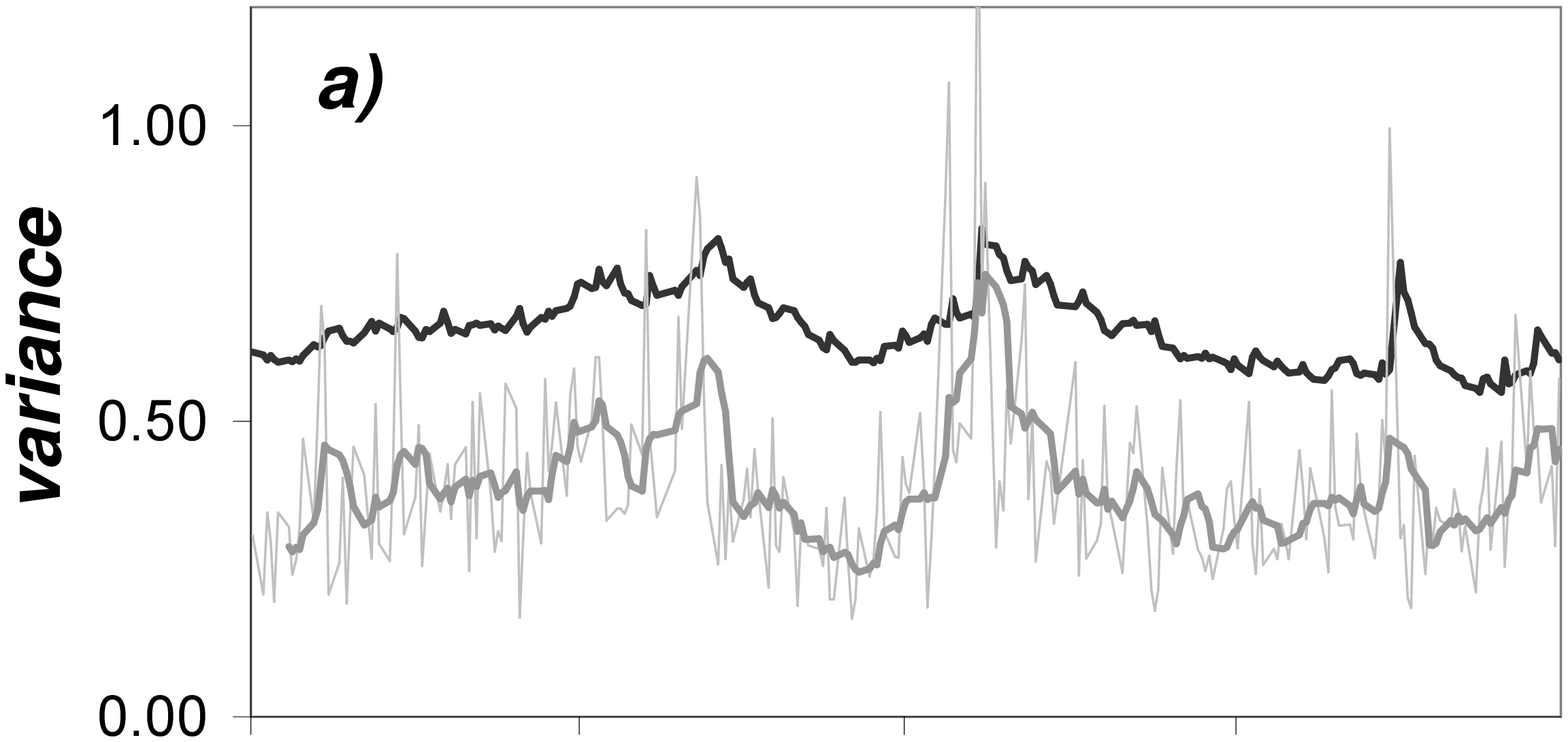,width=8cm}}
\vspace{-.8cm}
\centerline{ \epsfig{file=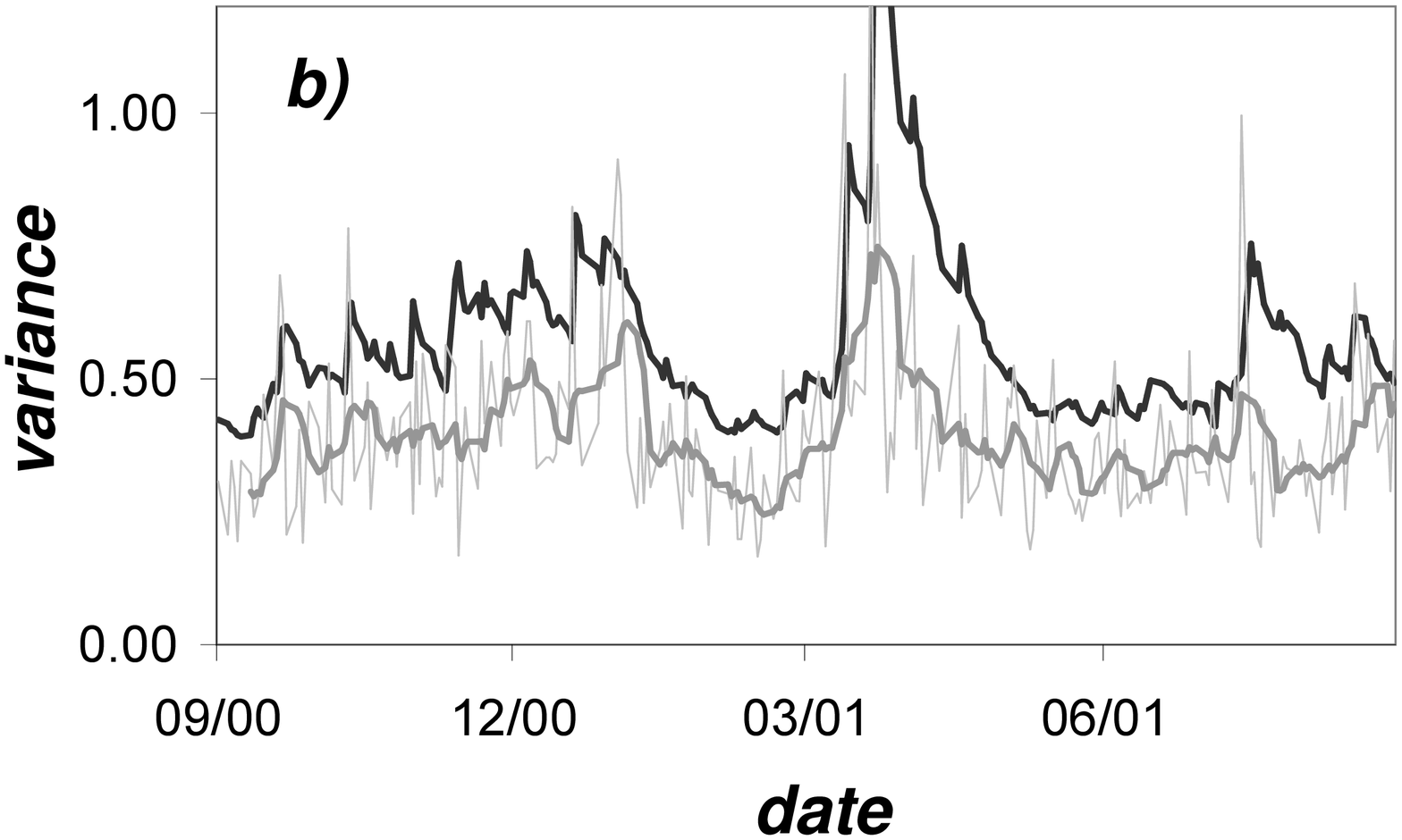,width=8cm}}
\caption{
  Prediction of daily volatility of a), b) equally weighted portfolios
  and c), d) minimum variance portfolios.  The black line shows the
  estimated volatility, the thin grey line the realized volatility
  calculated from high frequency data while the thick grey line is an
  seven-day average of the realized volatility.  a) Sliding
  correlation model SCM(1)-seven, which uses one principal component
  and is estimated over a seven--year period, overestimates the
  variance by 65 \% due to an overestimation of the average
  correlation strength.  b) Model Factor(1)-seven, which uses one
  factor and is estimated over a seven--year period is able to
  describe changes in the correlation strength but tends to
  overestimate correlations in the time after a volatility peak.
    c) RiskMetrics estimator Eq. \ref{riskmetrics.eq} predicts a very
  small variance for a minimum variance portfolio, while the realized
  variance is a factor 35 larger.  d) Sliding correlation
  model SCM(1)-one uses one principal component and estimates
  correlations over a sliding time window of one year.  The predicted
  variance is close to the realized one, which is more than 50 \%
  smaller than for the RiskMetrics estimator. }
\label{mean.fig}
\end{figure}

\noindent
{\bf Test method}

\noindent
We judge the quality of a multivariate volatility model by its ability
to forecast the daily variance of both equally distributed and minimum
variance portfolios. The predicted portfolio volatility at time $t$ is
given by
%
\begin{equation}
    D^2_{t,{\rm predicted}} = \sum_{i,j=1}^N m_{i,t} \; m_{j,t} \; 
\Sigma_{ij,t} \  \ .
\end{equation}
%
Here, $m_{i,t}$ is the fraction of the capital invested in
stock $i$ at time $t$.  For an equally distributed portfolio, we have $m_{i,t}
\equiv 1/N$, hence $D^2$ is just the average element of the covariance
matrix.  This test probes the ability of a model to correctly predict
the average covariance between pairs of stock.  On the other hand,
%
\begin{equation}
m_{i,t} = \sum_{j=1}^N  \Sigma^{-1}_{ij,t}\, {\large /} \, \sum_{k,l=1}^N
\Sigma^{-1}_{kl,t}\ 
\end{equation}
%
minimize the variance under the constraint that the total invested
money is equal to one. The volatility prediction for minimum variance
portfolios is a difficult test for volatility models, as the
covariance matrix is first used to calculate the portfolio weights and
then to estimate the variance of that portfolio. We compare the
predicted volatility to the realized volatility, which is calculated
from high frequency (hourly) data \cite{Anderson97}. The realized
portfolio variance at time $t$ is calculated as
%
\begin{equation}
D^2_{t,{\rm realized}}= \sum_{i,j=1}^N m_{i,t} m_{j,t}
\left(\langle g_{i,t} g_{j,t}\rangle  - \langle g_{i,t}\rangle
\langle g_{j,t}\rangle\right)
\end{equation}
%
where the expectation values are taken for 10 hourly returns $g_i$ per
trading day. We perform an out of sample test from 09/01/00 until
08/31/01.  We calculated the average predicted portfolio
variance, the average realized variance, the mean square error (MSE)
of the prediction and the mean average percentage error (MAPE).\\

\noindent{\bf Test results:}

\begin{figure}[t]
\centerline{ \epsfig{file=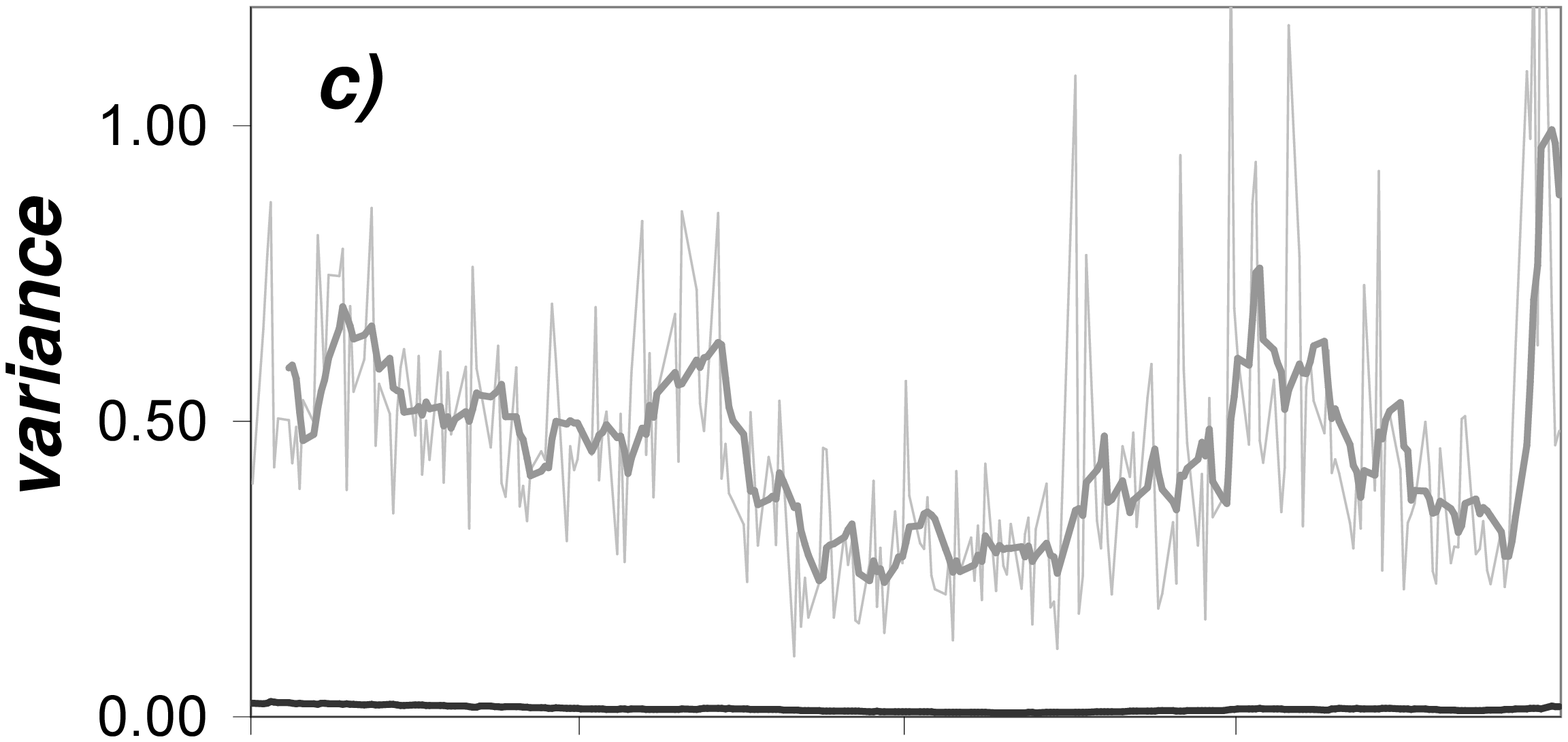,width=8cm}}
\vspace{-.8cm} 
\centerline{
\epsfig{file=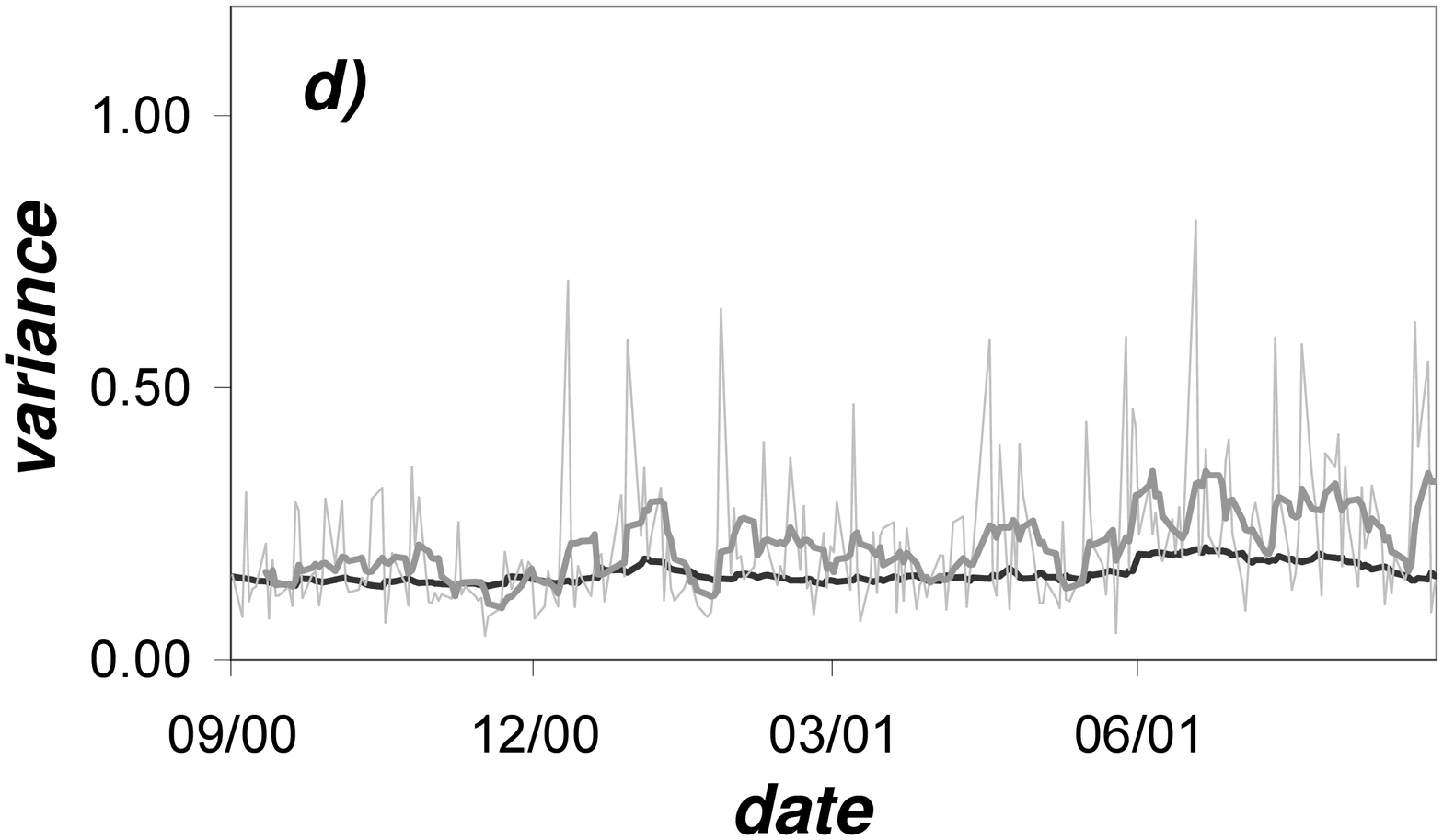,width=8cm}}
\end{figure}

\noindent
For {\it equally distributed portfolios}, the performance of the
various factor models is very similar to each other.  Model
Factor(1)-seven, for example, overestimates the overall level of
variance but is able to describe the effect of enhanced correlations
in high volatility periods (Fig. \ref{mean.fig}).

The variance estimation of SCM(N)-seven and SCM(1)-seven is very
similar to each other, both models severely overestimate the average
level of correlations, see Fig.  \ref{mean.fig} for the results of
SCM(1)-seven.  The sliding correlation models with a one year time
window for correlation estimates provide for the best estimates of
realized variances, the RiskMetrics estimator shows comparable
results. These results indicate that a good multivariate volatility
model must be able to capture not only the change of volatility over
time but also the change of the average correlation strength. For
details see Table \ref{table1.tab}.

\begin{table}[t]
\begin{tabular}{|c|c|c|c|c|} \hline
Model  &     $D^2_{\rm estimated}$ & $D^2_{\rm realized}$  & MSE & MAPE \\ \hline
SCM(1)-seven     & 0.654 & 0.396 & 0.092 & 0.886 \\ \hline
SCM(2)-seven     & 0.666 & 0.396 & 0.098 & 0.916 \\ \hline
SCM(N)-seven     & 0.643 & 0.396 & 0.086 & 0.856 \\ \hline
SCM(1)-one       & 0.497 & 0.396 & 0.037 & 0.486 \\ \hline
SCM(2)-one       & 0.500 & 0.396 & 0.037 & 0.493 \\ \hline
SCM(N)-one       & 0.484 & 0.396 & 0.034 & 0.456 \\ \hline\hline
Factor(1)-seven  & 0.568 & 0.396 & 0.063 & 0.615 \\ \hline
Factor(2)-seven  & 0.576 & 0.396 & 0.065 & 0.634 \\ \hline
Factor(1)-one    & 0.571 & 0.396 & 0.062 & 0.618 \\ \hline
Factor(2)-one    & 0.575 & 0.396 & 0.064 & 0.629 \\ \hline\hline
RiskMetrics      & 0.519 & 0.396 & 0.057 & 0.535 \\ \hline
\end{tabular}
\caption{Estimation of the variance of equally distributed portfolios}
\label{table1.tab}
\end{table}

\begin{table}[t]
\begin{tabular}{|c|c|c|c|c|} \hline
Model  &     $D^2_{\rm estimated}$ & $D^2_{\rm realized}$  & MSE & MAPE \\ \hline
SCM(1)-seven     & 0.177 & 0.243 & 0.022 & 0.335 \\ \hline
SCM(2)-seven     & 0.188 & 0.238 & 0.025 & 0.415 \\ \hline
SCM(N)-seven     & 0.190 & 0.245 & 0.029 & 0.409 \\ \hline
SCM(1)-one       & 0.158 & 0.208 & 0.016 & 0.340 \\ \hline
SCM(2)-one       & 0.157 & 0.208 & 0.017 & 0.338 \\ \hline
SCM(N)-one       & 0.128 & 0.284 & 0.046 & 0.479 \\ \hline\hline
Factor(1)-seven  & 0.215 & 0.304 & 0.033 & 0.318 \\ \hline
Factor(2)-seven  & 0.238 & 0.303 & 0.035 & 0.375 \\ \hline
Factor(1)-one    & 0.212 & 0.300 & 0.031 & 0.319 \\ \hline
Factor(2)-one    & 0.220 & 0.312 & 0.033 & 0.305 \\ \hline\hline
RiskMetrics      & 0.013 & 0.453 & 0.241 & 0.967 \\ \hline
\end{tabular}
\caption{Estimation of the variance of minimum variance
portfolios} \label{table2.tab}
\vspace*{-.6cm}
\end{table}

{\it Minimum variance portfolio:} Due to the measurement noise
contained in its cross--correlation matrix, the model SCM(N)-one
clearly underestimates the variance of minimum variance portfolios,
the ratio of realized variance to predicted variance is 2.22.  The
filtered model SCM(1)-one performs much better in this respect, the
ratio is given by 1.32 (Fig. 2d). In addition, the realized variance
of the ``filtered'' model is 27\% lower than that of the original
model.  Adding one more eigenvalue in the cross--correlation matrix
does not change this result much.  The performance of the various
factor models with one or two factors and estimation periods of seven
or one year is quite similar to each other, therefore we discuss the
representative model Factor(1)-seven (detailed results can be found in
Table \ref{table2.tab}).  Although the ratio of realized to predicted
variance is with 1.46 better than that of model SCM(N)-one, the
realized variance itself is higher than that of the other models.
Possibly the performance of the factor models could be increased if
the factor loadings were estimated by the maximum likelihood method
instead of using the regression coefficients.  Details of the results
can be found in Table \ref{table2.tab}. All these results compare
favorably to the RiskMetrics covariance estimator, which
underestimates the portfolio variance by a factor of 35 (Fig. 2c).
Also, the results reported in \cite{Engle+01} for DCC-GARCH
(overestimation of risk by several hundred percent) are much less
precise than the estimates of SCM(1)-one, for example.

In summary, we have shown that the ``curse of dimensionality'' can be
overcome by using appropriate statistical tools relying on RMT.
Building on the distinction between statistically relevant information
and noise in the correlation matrix, we have suggested two types of
multivariate GARCH--models that are easy to estimate even for more
than 100 time series.  Both the sliding correlation model and the
factor model studied have a much better prediction quality than the
industry standard RiskMetrics covariance estimator.

\end{document}